# Spinodal decomposition, nuclear fog and two characteristic volumes in thermal multifragmentation

V.A.Karnaukhov[1,*], H.Oeschler[2], S.P.Avdeyev[1], V.K.Rodionov[1], V.V.Kirakosyan[1],
A.V.Simonenko[1], P.A.Rukoyatkin[1], A.Budzanowski[3], W.Karcz[3], I.Skwirczyńska[3],
E.A.Kuzmin[4], L.V.Chulkov[4], E.Norbeck[5], A.S.Botvina[6]

[1] Joint Institute for Nuclear Research, 141980 Dubna, Russia
[2] Institut für Kernphysik, Darmstadt University of Technology, 64289 Darmstadt, Germany
[3] H.Niewodniczanski Institute of Nuclear Physics, 31-342 Cracow, Poland
[4] Kurchatov Institute, 123182 Moscow, Russia
[5] University of Iowa, Iowa City, IA 52242, USA
[6] Institute for Nuclear Research, 117312 Moscow, Russia

Thermal multifragmentation of hot nuclei is interpreted as the nuclear *liquid-fog* phase transition inside the spinodal region. The experimental data for $p(8.1\text{GeV}) + \text{Au}$ collisions are analyzed within the framework of the statistical multifragmentation model (SMM) for the events with emission of at least two IMFs. It is found that the partition of hot nuclei is specified after expansion to a volume equal to $V_t = (2.6 \pm 0.3)\, V_o$, with $V_o$ as the volume at normal density. However, the freeze-out volume is found to be twice as large: $V_f = (5 \pm 1)\, V_o$.

## 1. THERMAL MULTIFRAGMENTATION AND NUCLEAR FOG

The study of the decay of very excited nuclei is one of the most challenging topics of nuclear physics giving access to the nuclear equation of state for the temperatures below $T_c$ – the critical temperature for the liquid-gas phase transition. The main decay mode of very excited nuclei is a copious emission of intermediate mass fragments (IMF), which are heavier than α-particles but lighter than fission fragments. The great activity in this field has been stimulated by the expectation that this process is related to a phase transition in nuclear media.

An effective way to produce hot nuclei is reactions induced by heavy ions with energies up to hundreds of MeV per nucleon. But in this case the heating of the nuclei is accompanied by compression, strong rotation, and shape distortion, which may essentially influence the decay properties of hot nuclei. One gains simplicity, and the picture becomes clearer, when light relativistic projectiles (protons, antiprotons, pions) are used. In this case, fragments are emitted by only one source – the slowly moving target spectator. Its excitation energy is almost entirely thermal. Light relativistic projectiles provide therefore a unique possibility for investigating *thermal multifragmentation*. The decay properties of hot nuclei are well described by statistical models of multifragmentation [1, 2] and this can be considered as an indication that the system is thermally equilibrated or, at least, close to that. For the case of peripheral heavy ion collisions the partition of the excited system is also governed by heating.

In several papers, multifragmentation of hot nuclei is considered as spinodal decomposition. The appearance of the unstable spinodal region in the phase diagram of the nucleonic system is a consequence of the similarity between nucleon-nucleon and van der

[*]Email address: karna@jinr.ru

Waals interactions [3-5]. The equations of state are similar for these systems, which are very different in respect to the size and energy scales.

One can imagine that a hot nucleus (at $T = 7$-$10$ MeV) expands due to thermal pressure and enters the unstable region. Due to density fluctuations, a homogeneous system is converted into a mixed phase consisting of droplets (IMF) and nuclear gas interspersed between the fragments. Thus the final state of this transition is a *nuclear fog* [5], which explodes due to Coulomb repulsion and is detected as multifragmentation. It is more appropriate to associate the spinodal decomposition with the *liquid-fog* phase transition in a nuclear system rather than with the *liquid-gas* transition, as stated in several papers (see for example [6-8]).

This scenario is evidenced by number of the observations, some of them are the following:
(a) density of the system at the break-up is much lower compared to the normal one $\rho_o$;
(b) mean life-time of the fragmenting system is very small ($\approx 50$ *fm/c*), which is in the order of the time scale of density fluctuation [9];
(c) break-up temperature is lower than the critical temperature for the *liquid-gas* phase transition, which is found to be $T_c = (17 \pm 2)$ MeV [10].

The first point from this list requires more detailed experimental study. There are a number of papers with estimates of the characteristic volume (or mean density), but the values obtained deviate significantly. A mean freeze-out density of about $\rho_o/7$ was found in Ref. [11] from the average relative velocities of the IMFs at large correlation angles for $^4$He(14.6 MeV) + Au collisions using the statistical model MMMC [2]. In paper [12] the nuclear caloric curves were considered within an expanding Fermi gas model to extract average nuclear densities for different fragmenting systems. It was found to be $\sim 0.4$ $\rho_o$ for medium and heavy masses. In Ref. [13] the mean kinetic energies of fragments were analyzed by applying energy balance, calorimetric measurements and Coulomb trajectories calculations. The freeze-out volume was found to be $\sim 3 V_o$ for the fragmentation in Au(35·A MeV) + Au collisions. The average source density for the fragmentation in the 8.0 GeV/$c$ $\pi^-$ + Au interaction was estimated to be $\sim (0.25$-$0.30)\rho_o$ at $E^*/A \sim 5$ MeV from the moving-source-fit Coulomb parameters [7].

In our paper [14], the inclusive data on the charge distribution and kinetic energy spectra of IMFs produced in $p$(8.1 GeV) + Au collisions were analyzed using the statistical model SMM [1]. It was concluded that one should use *two* volume (or density) parameters to describe the multifragmentation process. The first, $V_t = (2.9 \pm 0.2) V_o$, corresponds to the stage of fragment formation, the second $V_f = (11 \pm 3) V_o$, is the freeze-out volume.

In the present paper we analyze data for the events with fragment multiplicity $M$ of two or more. The conclusion about two characteristic volumes is confirmed, but the value of the freeze-out volume is found to be less than the one obtained for the inclusive data: $V_f (M \geq 2) = (5 \pm 1) V_o$. A possible reason for this difference is discussed. Experimental data have been obtained using the 4$\pi$-device FASA installed at the external beam of the Nuclotron (Dubna) [15]. This setup consists of five *dE-E* telescopes surrounded by a fragment multiplicity detector (FMD), which is composed by 64 scintillation counters with thin CsI(Tl).

## 2. VOLUME FROM IMF CHARGE DISTRIBUTIONS

The reaction mechanism for light relativistic projectiles is usually divided into two stages. The first is a fast energy-depositing stage during which very energetic light particles are emitted and the target spectator is excited. We use the intranuclear cascade model (INC) [16] for describing the first stage. The second stage is considered within the framework of the SMM, which describes multibody decay (volume emission) of a hot and expanded nucleus. But such a two-stage approximation fails to predict the measured fragment multiplicity. To overcome this difficulty, an expansion stage is inserted (in the spirit of the EES model [17]). The residual (after INC) masses and their excitation energies are tuned (on event-by-event basis) to obtain agreement with the measured mean IMF multiplicity [18]. We call this combined model the INC+Exp+SMM approach.

The break-up (or partition) volume is parameterized in the SMM as $V = (1+k) V_o$. It is assumed in the model that the freeze-out volume, defining the total Coulomb energy of the final channel, coincides in size with the system volume when the partition is specified. Thus, $k$ is the only volume parameter of the SMM, which also defines (in first approximation) the free volume ($\approx kV_o$) and the contribution of the translation motion of the fragments to the entropy of the final state. Within this model the probabilities of different decay channels are proportional to their statistical weights (exponentials of entropy). The entropy is calculated using the liquid-drop model for hot fragments. The statistical model considers the secondary disintegration of the excited fragments to get the final charge distribution of cold IMFs. The importance of the secondary decay stage is analyzed in Ref. [9].

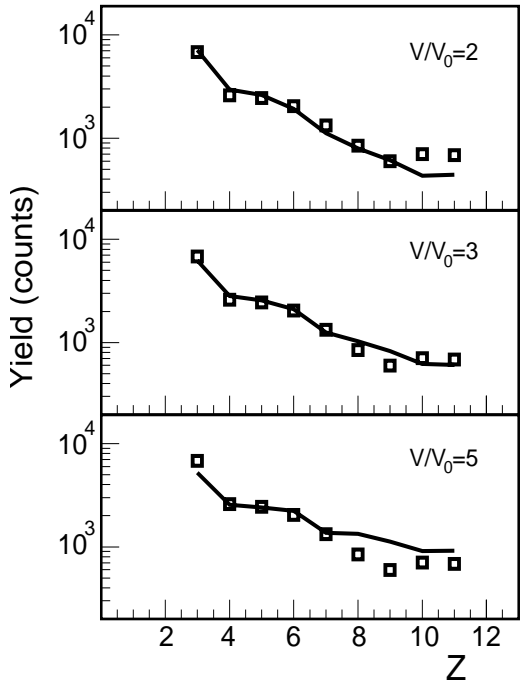
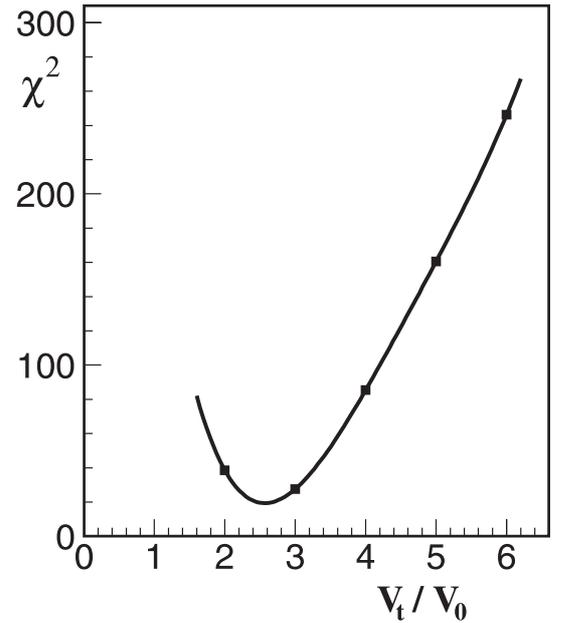

Fig.1. Charge distributions of intermediate mass fragments measured for $p$(8.1GeV) + Au collisions (dots) and calculated with the INC+Exp+SMM prescription using different values of the system volume, $V_t$, at the stage of fragment formation.

Fig.2. Value of $\chi^2$ as a function of $V_t/V_o$ for comparison of the measured and calculated IMF charge distributions. The best fit of the model prediction to the data corresponds to $V_t = (2.6 \pm 0.3)V_o$.

Figure 1 shows the IMF charge distribution for $p$ (8.1GeV) + Au collisions measured by a telescope at $\theta = 87^o$, provided that at least one more IMF is detected by FMD. Error bars do not exceed the symbol size. The lines are obtained by calculations using the INC+Exp+SMM prescription under three assumptions about the fragmenting system volume: $2V_o$, $3V_o$ and $5V_o$. Only events with IMF multiplicity $M \geq 2$ are considered. The experimental filter of FMD has been taken into account. The theoretical charge distributions are normalized to get the total fragment yield equal to the measured one in the $Z$ range between 3 and 11. A remarkable density dependence of the calculated charge distributions is visible.

The least-square method has been used for quantitative comparison of the data and the calculations. Figure 2 shows the normalized $\chi^2$ as a function of $V/V_o$. From the minimum position and from the shape of the curve in its vicinity it is concluded that the best fit is obtained with the partition volume $V_t = (2.6 \pm 0.3) V_o$. The error bar ($2\sigma$) is statistical in origin. This value corresponds to a mean density of the system $\rho_t = (0.38 \pm 0.04)\rho_o$ (see later why the subscript " t " is used).

## 3. SIZE OF EMITTING SOURCE

Generally, the fragment kinetic energy is determined by thermal motion, Coulomb repulsion, rotation, and collective expansion, $E = E_{th} + E_C + E_{rot} + E_{flow}$. The Coulomb term is about three times larger than the thermal one [9]. The contributions of the rotational and flow energies are negligible for $p$+Au collisions [18]. So, the energy spectrum is essentially sensitive to the size of the emitting source. The kinetic energy spectra are obtained by calculation of multibody Coulomb trajectories, which starts with placing all charged particles of a given decay channel inside the freeze-out volume $V_f$. Each particle is assigned a thermal momentum corresponding to the channel temperature. The Coulomb trajectory calculations are performed for 3000 fm/$c$. After that the fragment kinetic energies are close to the asymptotic values [9]. These calculations are the final step of the INC+Exp+SMM combined model.

We analyzed carbon spectrum measured by a telescope at $\theta = 87^o$ under the condition that at least one additional IMF is detected by FMD. Figure 3 gives a comparison of the measured spectrum with the calculated ones (for emission polar angles $\theta = 87^o \pm 7^o$). Calculations have been done for the events with $M_{IMF} \geq 2$ taking into account the experimental filter of FMD. The energy ranges of the spectra are restricted to 80 MeV to exclude the possible contribution of preequilibrium emission. The calculations are performed with a fixed partition volume, $V_t = 2.6\ V_o$, in accordance with the findings of the previous section. The freeze-out volume, $V_f$, is taken as a free parameter. Figure 3 shows the calculated spectra for $V_f / V_o$ equals to 3, 6 and 13. The least-square method is used to find the value of $V_f$ corresponding to the best description of the data. Figure 4 presents $\chi^2$ as a function of $V_f / V_o$. From the position of its minimum one gets $V_f = (5 \pm 1) V_o$ (or mean freeze-out density $\rho_f \approx 0.2\ \rho_o$). Systematics provides the main contribution to the error of this estimation of the freeze-out volume. It is caused by a 5% uncertainty in the energy scale calibration. In our recent paper [14] the value $V_f = (11 \pm 3) V_o$ was obtained by analyzing the inclusive energy spectrum of carbon. This great difference may be explained by the fact that SMM overestimates fragment energies for the events with $M = 1$ [19]. As result, the fitting procedure [14] shifts $V_f$ to the larger values.

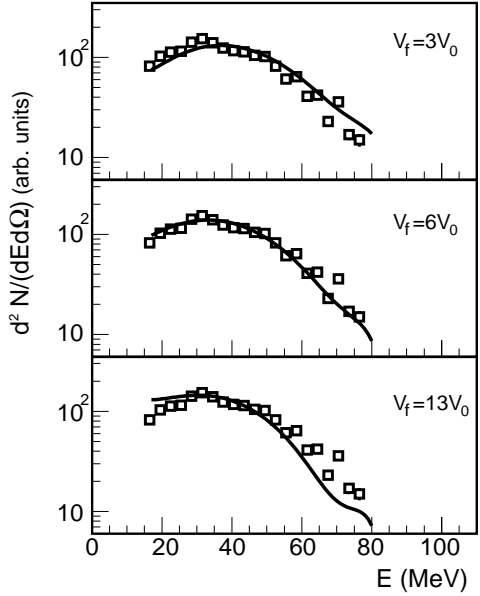
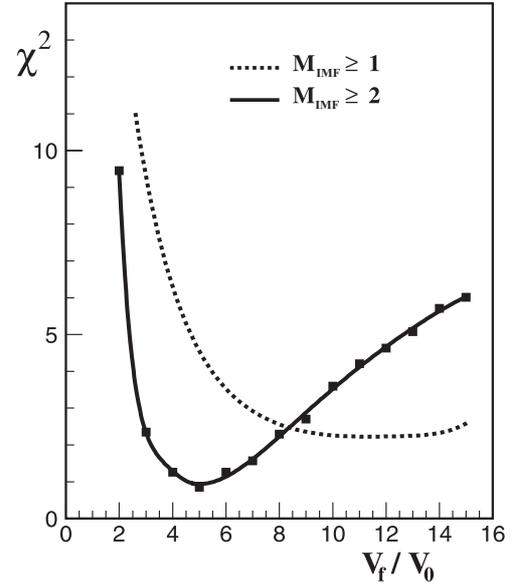

Fig.3. Kinetic energy spectrum of carbon emitted (at $\theta=87^{o}$) by the target spectator in $p(8.1\text{GeV}) + Au$ collisions. Symbols are the data, lines are calculated assuming $V_t = 2.6V_o$. The freeze-out volume, $V_f$, is taken to be equal to 3, 6, and 13 $V_o$ (upper, middle, bottom panels).

Fig.4. Value of $\chi^2$ as a function of the freeze-out volume $V_f/V_o$ for comparison of the measured and calculated kinetic energy spectra of carbon. The solid line is for the events with IMF multiplicity $M \geq 2$. The best fit corresponds to $V_f = (5 \pm 1) \ V_o$. Dashed line is for the inclusive data [14].

## 4. DISCUSSION

Figure 5 presents the proposed spinodal region in the $T-\rho$ plane [3] with the experimental data obtained in present paper. The points for the partition and freeze-out configurations are located at $\rho_t$ and $\rho_f$. Corresponding temperatures have been determined by fitting the data for fragment yields with the statistical model calculations [9, 8]. The value of the fragmentation barrier has been taken into account. The two points are deep inside the spinodal region, the top of which is specified by the critical temperature for the liquid-gas phase transition [10].

The existence of two different size characteristics for multifragmentation has a transparent meaning. The first volume, $V_t$, corresponds to the partition point (or the moment of fragment formation), when the properly extended hot target spectator transforms into a configuration consisting of specified prefragments. They are not yet fully developed, there are still links (nuclear interaction) between them. The final channel of disintegration is completed during the evolution of the system up to the moment when receding and interacting prefragments become completely separated. This is just as in ordinary fission. The saddle point (which has a rather compact shape) resembles the final channel of fission by way of having a fairly well-defined mass asymmetry. Nuclear interaction between fission prefragments cease after descent of the system from the top of the barrier to the scission point. In papers by Lopez and Randrup [20] the similarity of both processes was used to develop a theory of multifragmentation based on suitable generalization of the transition-state approximation first considered by Bohr and Wheeler for ordinary fission. The theory is able to calculate the potential energy as a function of the rms. extension of the system yielding the space and energy characteristics of the transition configuration and the barrier height for fragmentation. The transition state is located at the top of the barrier or close to it. The phase space properties of the transition state are decisive for its further fate, for specifying the final channel.

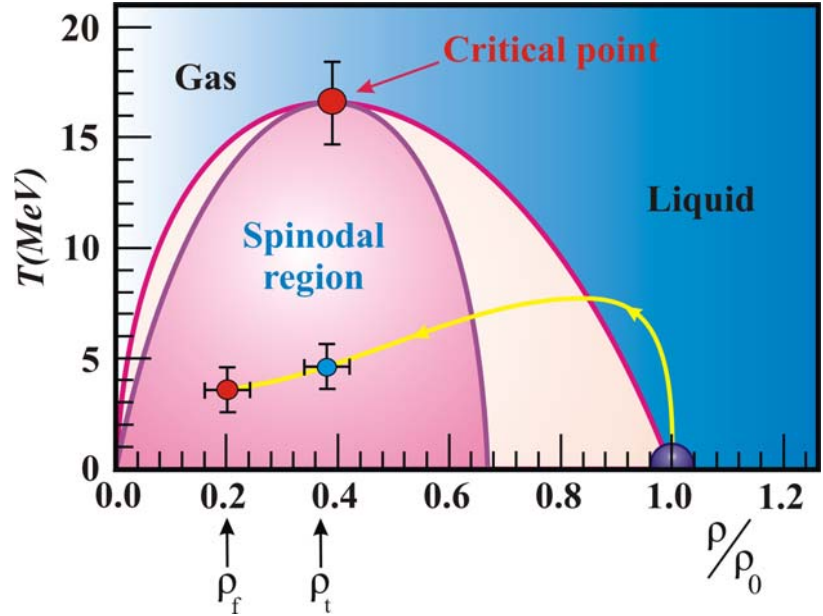

Fig.5. Proposed spinodal region for nuclear system. The experimental points were obtained by the FASA collaboration. The arrow line shows the way of the system from the starting point at $T=0$ and $\rho_o$ to the multi-scission point at $\rho_f$.

Being conceptually similar to the approach of Ref. [20], the statistical model of multifragmentation (SMM) uses the size parameter, which can be determined by fitting to data. The size parameter obtained from the IMF charge distribution can hardly be called a freeze-out volume. In the spirit of the papers by Lopez and Randrup we suggest the term "transition state volume", $V_t = (2.6 \pm 0.3)\, V_o$.

The larger value of the size parameter obtained by the analysis of the kinetic energy spectra is a consequence of the main contribution of Coulomb repulsion to the IMF energy, which starts to work when the system has passed the " multi-scission point " (see Fig. 6). Thus, $V_f = (5 \pm 1)\, V_o$ is the freeze-out volume for multifragmentation in $p + Au$ collisions. It means that the nuclear interaction between fragments is still significant when the system volume is equal to $V_t$, and only when the system has expanded up to $V_f$, are the fragments freezing out. In the statistical model used, the yield of a given final channel is proportional to the corresponding statistical weight. Therefore, the nuclear interaction between prefragments is neglected when the system volume is $V_t$, and this approach can be viewed as a rather simplified transition-state approximation. Nevertheless, the SMM describes well the IMF charge (mass) distributions for thermally driven multifragmentation. Note that in the traditional application of the SMM only one size parameter is used, which is called "freeze-out volume". In the present paper we have demonstrated the shortcoming of such a simplification of the model.

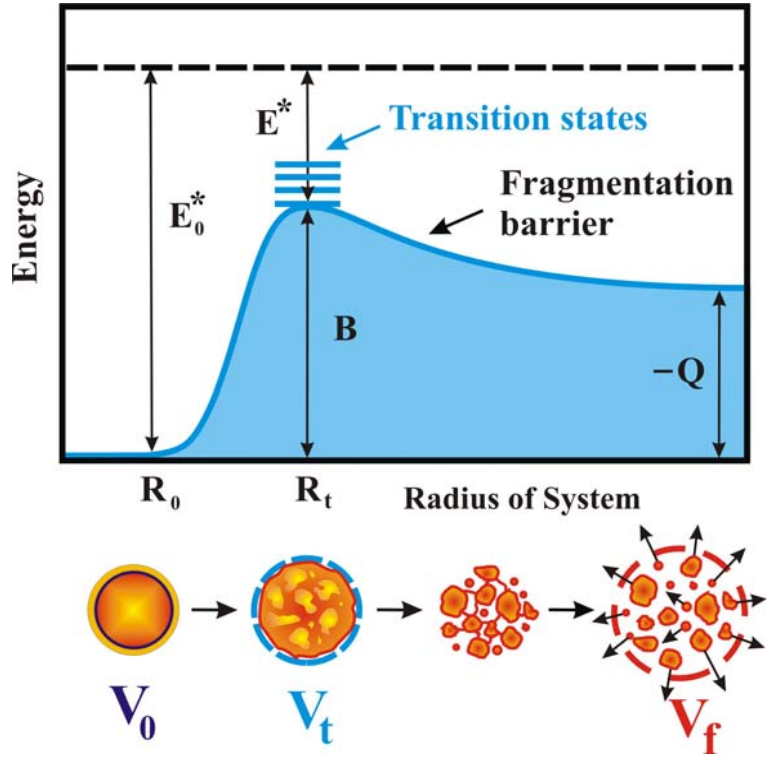

Fig.6. Upper: qualitative presentation of the potential energy of the hot nucleus (with excitation energy $E_o^*$) as a function of the system radius. Ground state energy of the system corresponds to $E=0$, $B$ is the fragmentation barrier, $Q$ is the released energy. Bottom: schematic view of the multifragmentation process.

The values of $V_t$ and $V_f$ may be sensitive to the way of their estimation. One could imagine that the freeze-out volume $V_f$ might be estimated by a model independent method if the experimental data on the source $Z$ value and charge distribution in the final channel were known. After that one needs to calculate the multi-body Coulomb trajectory (with $V_f$ as a single free parameter) to get fragment energies. But it is true only for the case when the fragments are already cold after scission point as in MMMC model [2]. As for $V_t$, we do not see any possibility of finding it in a model independent way. We know that it is a key parameter for defining the fragment charge distribution. But one should look for other observables that are sensitive to the $V_t$ value.

In a recent paper by Campi et al. [21] the "little big bang" (LBB) scenario of multifragmentation is suggested in which fragments are produced at an early, high temperature and high density stage of the reaction (in contrast to the statistical models). This scenario is very impressive, but it is not yet a well-finished model that could be compared directly with experimental data. We are looking forward to analysing our data with the well-developed LBB approach.

## 5. SUMMARY

Analysis of the data for fragmentation in $p(8.1\text{GeV}) + \text{Au}$ collisions (for events with $M_{\text{IMF}} \geq 2$) results in the conclusion that within the framework of the SMM there are two characteristic volume (or density) parameters. One, $V_t = (2.6 \pm 0.3) V_o$, is obtained from the IMF charge distribution. It corresponds to the configuration of the system at the stage of prefragment formation. It is similar to the saddle point in ordinary fission (transition state). The other, $V_f = (5 \pm 1) V_o$, is found from the analysis of the fragment energy spectra. It is the freeze-out volume corresponding to the multi-scission point in terms of ordinary fission. In further studies of the size or density parameters it is important to specify which stage of the system evolution is relevant to the observable chosen for the analysis. Spinodal decomposition of hot nuclei is interpreted as a liquid-fog phase transition with the final state consisting of droplets (IMFs) with nuclear gas interspersed between them.

The authors are grateful to A. Hrynkiewicz, A.I. Malakhov, A.G. Olchevsky for support and to I.N. Mishustin and W. Trautmann for illuminating discussions. The research was supported in part by the Russian Foundation for Basic Research, Grant № 03-02-17263, the Grant of the Polish Plenipotentiary to JINR, Bundesministerium für Forschung und Technologie, Contract № 06DA453, and the US National Science Foundation.

**REFERENCES**

[1]  J.P. Bondorf, A.S. Botvina, I.N. Mishustin et al., Phys. Rep. **257** (1995) 133.
[2]  D.H.E. Gross, Rep. Progr. Phys. **53** (1990) 605.
[3]  G. Sauer, H. Chandra and U. Mosel, Nucl. Phys. **A264** (1976) 221.
[4]  H. Jaqaman, A.Z. Mekjian and L. Zamick, Phys. Rev. C **27** (1983) 2782.
[5]  P.J. Siemens, Nature **305** (1983) 410; Nucl. Phys. **A428** (1984) 189c.
[6]  K.A. Bugaev, M.I. Gorenstein, I.N. Mishustin and W. Greiner, Phys. Rev. C **62** (2000) 044320.
[7]  V.E. Viola, Nucl. Phys. **A734** (2004) 487.
[8]  B. Borderie, R. Bougault, P. Desesquelles et al., Nucl. Phys. **A734** (2004) 495.
[9]  V.K. Rodionov, S.P. Avdeyev, V.A. Karnaukhov et al., Nucl. Phys. **A700** (2002) 457.
[10] V.A. Karnaukhov, H. Oeschler, S.P. Avdeyev et al., Nucl. Phys. **A734** (2004) 520.
[11] Bao-An Li, D.H.E. Gross, V. Lips and H. Oeschler, Phys. Lett. B **335** (1994) 1.
[12] J.B. Natowitz, K. Hagel, Y. Ma et al., Phys. Rev. C **66** (2002) 031601(R).
[13] M.D. Agostino, R. Bougault, F. Gulminelli et al., Nucl. Phys. **A699** (2002) 795.
[14] V.A. Karnaukhov, H. Oeschler, S.P. Avdeyev et al., Phys. Rev. C **70** (2004) 1(R).
[15] S.P. Avdeyev, A.S. Zubkevich, V.A. Karnaukhov et al., Instr. Exp. Techn. **39** (1996) 153.
[16] V.D. Toneev, N.S. Amelin, K.K. Gudima and S.Yu. Sivoklokov, Nucl. Phys. **A519** (1990) 463.
[17] W.A. Friedman, Phys.Rev. Lett. **60** (1988) 2125.
[18] S.P Avdeyev, V.A. Karnaukhov, L.A. Petrov et al., Nucl.Phys. **A709** (2002) 392.
[19] H. Oeschler, A.S. Botvina, D.H.E. Gross et al, Particles and Nucl., Lett. №**2[99]** (2000) 70.
[20] (a)J.A. Lopez and J. Randrup, Nucl. Phys. **A503** (1989) 183; (b) Nucl. Phys. **A512** (1990) 345.
[21] X. Campi et al., Phys.Rev. C **67** (2003) 044610.